\documentclass[fleqn]{article}

\usepackage[sumlimits]{amsmath}
\usepackage{bbold}                    
\usepackage{amsfonts}
\usepackage{amssymb}
\usepackage{amsopn}
\usepackage{amsthm}                   
\usepackage{mathrsfs}                 
\usepackage{titlesec}
\usepackage[leftflush=9mm]{diagrams}
\usepackage{graphicx}
\usepackage{multicol}
\titleformat{\section}[hang]
 {\large\sffamily\bfseries}
 {\thesection}{.5em}{}
\titleformat{\subsection}[hang]
 {\normalfont\sffamily\bfseries}
 {\thesubsection}{.5em}{}
\titleformat{\subsubsection}[hang]
 {\normalfont\sffamily}
 {\thesubsubsection}{.5em}{}
\titleformat{\paragraph}[runin]
 {\normalfont\sffamily\bfseries}
 {}{.5em}{}

\newcounter{cnt_example}
\newcounter{cnt_exercise}
\newcounter{cnt_table}

\newtheorem*{definition*}{\sc Definition}
\newtheorem*{corollary*}{\sc Corollary}
\newtheorem*{theorem*}{\sc Theorem}
\newtheorem*{lemma*}{\sc Lemma}
\newtheorem*{proposition*}{\sc Proposition}
 {\refstepcounter{cnt_example}%
  \par\smallskip\nopagebreak\noindent{\it Example \arabic{cnt_example}: }%
 }%
 {\qed\par\smallskip}
\newenvironment{example*}[1]%
 {\par\medskip\noindent{\it Example {#1}:} }%
 {\qed\par\smallskip}
 {\par\smallskip\noindent\small{\it Remark.}}%
 {\par\smallskip}
 {\refstepcounter{cnt_exercise}\par\medskip\noindent{\it Exercise \thesection{}.\arabic{cnt_exercise}.}}%
 {\par}

\renewenvironment{figure}%
 {\par\medskip\refstepcounter{figure}\footnotesize\sffamily\selectfont}%
 {\par\medskip}
 {\par\medskip\refstepcounter{cnt_table}\footnotesize\sffamily\selectfont}%
 {\par\medskip}

\def\dif{\mathrm{d}}                  
\def\imu{\mathrm{i}}                  
\DeclareMathOperator{\id}{id}         
\DeclareMathOperator{\img}{img}       
\DeclareMathOperator{\spct}{sp}       
\newcommand{\e}[1]{\mathrm{e}^{#1}}   
\newcommand{\term}[1]{\textit{#1}}    
\newcommand{\const}{\mathrm{const}}   
\newcommand{\ie}{\textit{i.e.}}       
\newcommand{\eg}{\textit{e.g.}}       
\newcommand{\cf}{\textit{cf.}}        
\newcommand{\etal}{\textit{et al.}}   

\def\thalf{\tfrac{1}{2}}              
\def\implies{\Rightarrow}             
\def\idmatrix{\text{\textbb 1}}       
\def\Till{,\!...,}                    
\def\fN{\mathbb{N}}                   
\def\fZ{\mathbb{Z}}                   
\def\fR{\mathbb{R}}                   
\def\fC{\mathbb{C}}                   
\def\intprod
 {{\unitlength 1pt\begin{picture}(9,8)(-7,0)\line(0,1){6}\line(-1,0){6}\end{picture}}}
\newcommand\roplus
 {{\unitlength 1pt\begin{picture}(11,8)(0,0)\linethickness{.3pt}%
  \put(5.0,-.3){\line(0,1){5.4}}\put(2,0){$\ni$}%
 \end{picture}}}
\newcommand\loplus
 {{\unitlength 1pt\begin{picture}(11,8)(0,0)\linethickness{.3pt}%
  \put(5.3,-.3){\line(0,1){5.4}}\put(2,0){$\in$}%
 \end{picture}}}
\def\grSL{\mathrm{SL}}                

\renewcommand{\caption}[1] {\par{\footnotesize\sffamily Figure \thefigure{}: #1}}

\newcommand{\figcaption}[1]{\par\nopagebreak{\textbf{Figure \thefigure{}} #1}}

\def\refno#1         {\noindent{\footnotesize #1}\par\noindent\rule[5mm]{\textwidth}{0.1mm}\par}
\newenvironment{articlehead}%
 {\begin{quote}\setlength{\parskip}{0pt}\setlength{\parindent}{0pt}}%
 {\end{quote}}
\def\title#1         {{\sffamily\Large\selectfont\textbf{#1}}\\\par\medskip}
\renewcommand{\author}[1]{{\textsc{#1}}}
\def\authore#1#2     {\textsc{#1}\footnote{e-mail: {#2}}\null}

\def\authorae#1#2#3  {\textsc{#1}$^{#2,}$\footnote{e-mail: {#3}}\null}
\def\address#1       {\par{\it #1}}
\renewenvironment{abstract}%
 {\par\medskip\small\begin{center}\textbf{Abstract}\end{center}\par~~~}%
 {\par}
\def\keywords#1      {\par{\bf Keywords:~}{#1}\par}
\def\pacs#1          {\par{\bf PACS:~}{#1}\par}
\def\subjclass#1     {\par{\bf Subj. class.:~}{#1}\par}

\newcommand\TRASH[1]{}                

\def\Gamma{\varGamma}
\def\Theta{\varTheta}
\def\Lambda{\varLambda}
\def\Xi{\varXi}
\def\Pi{\varPi}
\def\Sigma{\varSigma}
\def\Upsilon{\varUpsilon}
\def\Phi{\varPhi}
\def\Psi{\varPsi}
\def\Omega{\varOmega}
\def\rho{\varrho}

\def\ODE{\textrm{\small ODE}}
\def\OdE{\textrm{\small O}$\Delta$\textrm{\small E}}
\def\PDE{\textrm{\small PDE}}
\def\IPM{{\rm\small IPM}}
\def\FP {\textrm{\small FP}}
\def\Poincare{Poincar\'{e}}

\DeclareMathOperator{\Acos}{Arccos}
\DeclareMathOperator{\acos}{arccos}
\DeclareMathOperator{\acosh}{arccosh}
\def\spX{\mathscr{M}}               

\def\spE{\mathscr{E}}
\def\stSym{\mathscr{S}}
\def\dimX{m}                        
\def\dimY{m}
\def\chX{U}                         

\def\x{x}                           
\def\y{y}
\def\vey{\vec{y}}
\def\ptx{x}
\def\pty{y}

\def\grX{\mathrm{G}}
\def\grY{\mathrm{G}_\mathrm{lin}}
\def\autX{\mathrm{H}}               
\def\autY{\mathrm{H}_\mathrm{lin}}
\def\greX{f}                        

\def\X{X}                           

\def\xS{\Phi}
\def\oX{\hat{X}}                    

\def\oxS{\hat{\Phi}}
\def\mpp{p}                         
\def\mps{\sigma}                    
\def\mpk{\kappa}                    
\def\mE{E}                          

\def\mL{L}                          

\def\mM{M}

\def\xsym{\phi}

\def\xsim{\phi}
\def\ysim{\mL}
\def\xSym{S}
\def\xaut{h}
\def\mpt{g}

\def\idi{i}
\def\idj{j}

\def\ibi{{\bar{i\,}\!}}

\def\ida{a}
\def\idb{b}
\def\idc{c}
\def\idd{d}

\def\iga{\alpha}
\def\igb{\beta}
\def\igc{\gamma}
\def\igd{\delta}

\begin{document}
\begin{articlehead}
\title{Symmetries and linearization of ordinary difference equations}
\author{P. Gralewicz}
\address{Department of Theoretical Physics, University of \L\'{o}d\'{z},
  Pomorska 149/153, 90-236, \L\'od\'z, Poland}
\begin{abstract}
  The connection between symmetries and linearizations of discrete-time dynamical
systems is being inverstigated. It is shown, that existence of semigroup
structures related to the vector field and having linear representations enables
reduction of linearization problem to a system of first order partial differential
equations.
  By means of inverse of the {\Poincare} map one can relate symmetries in such
linearizable systems to continuous and discrete ones of the corresponding
differential equations.
\end{abstract}
\end{articlehead}
\begin{multicols}{2}

\section{Introduction}
  The principal objective behind the concept of (exact) linearization is to
express an apparently involved evolution law in simple terms, facilitating
further analysis.
  The purpose is also to bring a system to some `canonical' form enabling
comparison among studied dynamical systems in terms of {\eg} minimal embedding
dimension, which can be regarded as a complexity measure.
  To that end one seeks a projection map $\mpp:\spE\to\spX$ from a suitable
linear space $\spE$ onto the phase space $\spX$, which defines a homomorphism of
appropriate (semi)group actions.
The desired feature is of course finiteness of the embedding space $\spE$,
otherwise the linearization remains formal.
On the other hand, it is worthwhile to recall the applications of linearization
in infinite-dimensional spaces. An example are the methods of Hilbert spaces in
spectral theory \cite{bb:SR75}.

  Since the works of Lie, the notion of a symmetry as an automorphism in the
space of solutions plays central role in the analysis of differential equations
\cite{bb:Olv86,bb:BlKu89,bb:Stp89}. In particular, it is well-known that the
knowledge of an infinitesimal Lie-point and generalized symmetriy, allows to
reduce the system of ordinary dyfferential equations ({\ODE}s).
  An analogous result for ordinary difference equations ({\OdE}s) was obtained
by Maeda \cite{bb:Mae87}, and expanded by Quispel {\etal} \cite{bb:QS93,bb:SBQ95,bb:SBQ96}.
Namely, they showed that determination of a Lie-point or generalized symmetry,
enables reduction of the dimension of an {\OdE}, and in particular cases its
linearization \cite{bb:Mae87,bb:SBQ95}. The approach taken up by Maeda and
Quispel {\etal} relies on finding an infinitesimal time-translation symmetry,
and transformation into normal coordinates. It should be noted, that
determination of such particular symmetry may be in general highly nontrivial
task.

  In this work, we introduce a method for the study of difference equations,
based upon observation that \emph{finite} symmetries of the evolution generator
form a structured set
of relations, identified as a semigroup of similarities. If at least one of
its closed subsets is known, then that structure can be used to form a
linear semigroup of isomorphic structure, and obtain differential system of
constraints for the desired projection map $\mpp$. This method is illustrated
by two examples: the `classical' Riccati equation related to the symmetry group
$\grSL(2)$, and a thorough analysis of a logistic equation,
with emphasis on the role played by similarities in classification of obtained
continuous-time solutions (the inverse {\Poincare} maps).

\section{Linearization and symmetries}
\label{sc:Sym}
Let us begin with a brief account of the problem of linearization of recurrences,
and its relation to the symmetries of evolution generators.

Let $\greX:\spX\to\spX$ be a function defined on a manifold $\spX$. Consider
the autonomous system of difference equations
\begin{equation}
 \label{eq:OdE}
  \x_{n+1} = f(\x_n),\qquad n\in\fN_0\equiv\fN\cup\{0\}.
\end{equation}
Iteration of the function $\greX$ generates an associated semigroup with
multiplication defined by composition of functions, and unity (monoid)
\[
  \grX := \{\greX^n \mid \greX^0=\id,\; \greX^{n+1}=\greX^n\circ\greX,\; n\in\fN_0 \}.
\]
  The \term{linearization} of Eq. \eqref{eq:OdE} is a pair $(\mL,\mpp)$
comprising a linear map $\mL:\spE\to\spE$ acting on a vector space $\spE$, and
projection $\mpp:\spE\to\spX$, such that the following diagram commutes:
\begin{diagram}[labelstyle=\scriptstyle,small,tight]
     \spE      & \rTo^{\mL^n}   & \spE      \\
     \dTo<\mpp &                & \dTo>\mpp \\
     \spX      & \rTo^{\greX^n} & \spX
\end{diagram}
for all $n\in\fN_0$. Equivalently, one can write
\begin{equation}
  \label{eq:defLin}
  \greX^n\circ\mpp = \mpp\circ\mL^n.
\end{equation}
Iterations of the linear map $\mL$ form another semigroup
\[
  \grY := \{\mL^n \mid n\in\fN_0\},
\]
which, acording to \eqref{eq:defLin} is homomorphic to $\grX$.
Therefore, the linearization ivolves
  (\textit{i}) fibration $\spE\overset{\mpp}{\to}\spX$,
 (\textit{ii}) linear function $\mL$, and
(\textit{iii}) semigroup homomorphism.

  Further, the kernel of $\mpp$ is a free monoid acting along the fibre:
\begin{equation}
  \label{eq:defKer}
  \begin{aligned}
  \ker\mpp
   &= \{\mpk:\spE\to\spE\mid\mpp\circ\mpk=\mpp\}
  \end{aligned}
\end{equation}
By definition, the elements of this kernel satisfy
\begin{equation}
  \label{eq:kLLk}
  \forall_{\mpk}\exists_{\mpk'}:\;
    \mpk\circ\mL = \mL\circ\mpk'
\end{equation}
where $\mpk,\mpk'\in\ker\mpp$. Notice, that the sections $\{\mps\}\in\mpp^{-1}$
and consequently the kernel elements too, are typically multivalued.
Whenever they appear in some expression, it is understood that the equality
holds true only for particular subset of these maps. For instance, the inverse
to Eq. \eqref{eq:defLin}
\[
  \mps\circ\greX^n = \mpk\circ\mL^n\circ\mps,
\]
means that for every $\mps\in\mpp^{-1}$ there exists $\mpk\in\ker\mpp$ (or
\textit{vice versa}) satisfying this relation.

  The iteration of function $\greX$ is often a result of temporal discretization,
and interpreted as the {\Poincare} map of some underlying, continuous-time
system. The reciprocal operation is what we call the \term{inverse {\Poincare}
map}\footnote{
  We use this term, since it ascribes certain physical meaning to the
  definition, however from mathematical viewpoint, a more apt name would be
  `monoidal homotopy' -- homotopy $f^t$ between $f^n$ and $f^{n+1}$ which
  additionally has the structure of a monoid.
} (\IPM), namely
an extension $(\greX^n,\spX)\mapsto(\greX^t,\tilde{\spX})$, such that
$\tilde{\spX}\supset\spX$, $\greX^t:\tilde{\spX}\to\tilde{\spX}$ for $t\in\fR_+$, and
\begin{enumerate}
\setlength{\itemsep}{0pt}\setlength{\parskip}{0pt}
  \item[(\textit{i})] $\greX^1\equiv\greX$,
  \item[(\textit{ii})] $\lim_{t\searrow 0}\greX^t=\id$,
  \item[(\textit{iii})] $\forall\,t,s\in\fR_+:\;\greX^t\circ\greX^s=\greX^{t+s}$,
\end{enumerate}
\noindent
In other words, the inverse {\Poincare} map is a one-parameter continuous
monoid $\{\greX^t\}$ congruent to $\{\greX^n\}$ at $t\in\fN_0$.
  The linearization $(\mL,\mpp)$ of Eq. \eqref{eq:OdE},
provides a suitable framework for obtaining {\IPM}s.
It must, however, be borne in mind that unlike $\mL^n$, the power $\mL^t$,
$t\in\fR_+$ is multivalued what makes characterization of the semigroup
$\grY$ by its generator $\mL$ imprecise.
One should instead use a particular branch of its logarithm $\mE$, that is
\[
  \mL^t=\exp(t\mE),
  \qquad
  \mL^t:\tilde{\spE}\to\tilde{\spE},\quad\spE\subset\tilde{\spE},
\]
which is well-defined. If the kernel of $\mpp$ is linear, then \eqref{eq:defKer}
implies $\mpk\circ\mE=\mE\circ\mpk'$.

  Further, the notion of inverse {\Poincare} map allows to discriminate between
apparently different linearizations, namely two linearizations $(\mpp_1,\mE_1)$,
$(\mpp_2,\mE_2)$ are regarded as \term{equivalent} if the corresponding {\IPM}s
\begin{align*}
  \greX_1^t &= \mpp_1\circ\e{t\mE_1}\circ\mps_1,\\
  \greX_2^t &= \mpp_2\circ\e{t\mE_2}\circ\mps_2,
\end{align*}
are homomorphic, {\ie}
\[
  \exists\,\xsim:\tilde{\spX}_1\rightrightarrows\tilde{\spX}_2:\;
    \xsim\circ\greX_1^t \cong \greX_2^t\circ\xsim,
\]
wwere the symbol `$\rightrightarrows$' indicates multivalence, and
`$\cong$' is used instead of `$=$', to emphasize that the branches of
map $\xsim$ on both sides may be different.

  According to the above definition, equivalence implies that, besides $\xsim$, there
must exist a map $U:\tilde{\spE}_1\to\tilde{\spE}_2$, such that
\begin{subequations}
 \label{eqs:equiv}
 \begin{align}
  &\mpp_2\circ U = \xsim\circ\mpp_1,\label{eq:pUup}\\
  &\mpk_2\circ\e{t\mE_2}\circ U = U\circ\e{t\mE_1}\circ\mpk_1.\label{eq:EUUE}
 \end{align}
\end{subequations}
For linear kernels the condition \eqref{eq:EUUE} reduces to `similarity'
between the logarithms $\mpk_2\circ\mE_2\circ U = U\circ\mE_1\circ\mpk_1$.
  Whenever the two flows are equal, $\greX_1^t=\greX_2^t$, the solution of
eqs. \eqref{eqs:equiv} is of the form
\begin{equation}
  \label{eq:U}
  U_\mathrm{eq} = \mps_2\circ\mpp_1,
\end{equation}
and $\xsim=\id$. Indeed, this reduces eq. \eqref{eq:pUup} to an identity, while
\eqref{eq:EUUE} can be verified by direct calculus:
\begin{align*}
  \mpk_2\circ \e{t\mE_2}\circ U_\mathrm{eq}
    &= \mps_2\circ\mpp_2\circ\e{t\mE_2}\circ \mps_2\circ\mpp_1\\
    &= \mps_2\circ\greX_2^t\circ\mpp_1\\
    &= \mps_2\circ\greX_1^t\circ\mpp_1\\
    &= \mps_2\circ\mpp_1\circ\mpk_1\circ\e{t\mE_1}\\
    &= U_\mathrm{eq} \circ \e{t\mE_1}\circ\mpk_1.
\end{align*}
If $\greX_1^t\not=\greX_2^t$, and one can find an alternative solution of
\eqref{eq:EUUE}, that is $U\not=U_\mathrm{eq}$, then
\begin{equation}
  \label{eq:xsim}
  \xsim=\mpp_2\circ U\circ\mps_1,
  \qquad
  \xsim:\spX\rightrightarrows\tilde{\spX}_2,
\end{equation}
is one of the desired homomorphism maps, establishing equivalence between these
two {\IPM}s.

  At the same time $\xsym$, becomes a \term{symmetry} of the function $\greX$
itself, or of its extension defined on $\tilde{\spX}_2$, that is
\begin{equation}
  \label{eq:defsym}
  \xsym\circ\greX \cong \greX\circ\xsym.
\end{equation}
Conversely, any such discrete symmetry provides similarity transformation
between two {\IPM}s.
If the image of $\xsym$ is $\spX$, then
it is a \term{proper} symmetry of the original function $f$.
Equation \eqref{eq:xsim} directly suggests another way of finding symmetries
by means of the linearization $(\mpp,\mL)$. Any linear mapping
$\mM:\tilde{\spE}\to\tilde{\spE}$ whose commutant with $\mL$ is an element of the kernel of
projection map
\begin{equation}
 \label{eq:ComtKer}
  \mL\circ\mM\circ\mL^{-1}\circ\mM^{-1}\in\ker\mpp,
\end{equation}
gives rise to a symmetry $\xsym$ defined by
\begin{equation}
 \label{eq:PhaseSym}
  \xsym:=\mpp\circ\mM\circ\mps.
\end{equation}
The case, $\mM=\mL^t$ for fixed $t\in\fR_+$ leads to the `trivial' symmetry
coinciding with the time-translation one $\xsym=\greX^t$.

The condition \eqref{eq:defsym}, can often be rewritten in alternative form as a
solution of certain implicit system
\[
  \xSym(\x,\xsym(\x))=0,
\]
such that the following implication holds
\begin{equation}
  \label{eq:defSym}
  \xSym(\ptx,\ptx')=0 \quad\implies\quad \xSym(\greX(\ptx),\greX(\ptx'))=0.
\end{equation}
Accordingly, the solutions of \eqref{eq:OdE} are invariant subsets of this
relation $\xSym(\x_n,\x_{n+1})\equiv 0$, {\ie} the symmetry maps solutions
into solutions.

  The symmetries $\phi(\x)$, similarly to linearizing transformations $\mpp$,
can be expanded locally into power series in $\x$ in the neighbourhood of fixed
points ({\FP}s) of the generating map $f$. Moreover, the same reasons which
cause nonexistence of (local) linear normal forms, also prevent unambiguous
symmetry expansion whenever $f$ has resonant spectra at at {\FP} $x_*$.
  If we denote by $x^i$, $i=1\Till\dim\spX$ the local coordinates of a point
$x\in\spX$, and the spectra of the first derivatives
$f^i\null\!_{,j}\equiv\partial f^i/\partial x^j$ by
\[
  \spct f^i\null\!_{,j}(x_*)=\{\lambda_{(1)}\Till\lambda_{(\dimX)}\},
  \qquad
  \dimX = \dim\spX,
\]
then this resonance condition takes the form
\[
  \exists\,\vec{k}\in\fN_0^\dimX:\;
  \prod_{i=1}^{\dim\spX}(\lambda_{(i)})^{k_i}\in \spct f^i\null\!_{,j}(x_*).
\]
If the condition \eqref{eq:defsym} is to be satified by a single branch of $\phi$
near {\FP} $x_*$, then one finds $\phi(x_*)=x_*$. The expansion
coefficients in the series 
\begin{equation}
 \label{eq:symexp}
  \phi(x_*+\delta{x})^i
    = x_*^i
    + \phi^i\null\!_{,j}\delta{x}^j
    + \thalf\phi^i\null\!_{,jk}\delta{x}^j \delta{x}^k
    + \ldots,
\end{equation}
where summation over repeated indices is understood,
are recursively defined functions of $f^i\null\!_{,j}(x_*)$ and
the free parameter tensor $\phi^i\null\!_{,j}(x_*)$.
In particular, chosing
\[
  \phi^i\null\!_{,j}(x_*)=(f^n)^i\null\!_{,j}(x_*), \qquad n\in\fZ,
\]
leads to the trivial symmetries $\phi = f^n$, $n\in\fN_0$. Nevertheless this
does not imply that $\phi$ coincides with the time translation symmetry $f^t$
for all $t\in\fR_+$, since in general $x_*$ is not a fixed point of $f^t$.
Other special
choices include $\phi^i\null\!_{,j}(x_*)=0\;\implies\;\phi\equiv x_*=\const$,
$\phi^i\null\!_{,j}(x_*)=\delta^i\null_j\;\implies\;\phi=\id$, and
$\phi^i\null\!_{,j}(x_*)=-\delta^i\null_j$ yields self-inverse symmetries,
{\ie} $\phi^{-1}=\phi$.
  In the second example provided in this article, we use local expansion
\eqref{eq:symexp}, together with the property \eqref{eq:defSym}, to investigate
numerically the symmetries of logistic map. In the same example, the many-valued
symmetries do not satisfy $\phi(x_*)=x_*$ for all fixed-points $\x_*$.

  Composition of symmetries form semigroups with identity (groups, if all
elements are invertible), which can be in general uncountable. The whole
set of symmetries of a given map $\greX$ is always non-empty, containing at
least the semigroup $\grX$ itself.
In particular, if a subset $\stSym:=\{\xsym^s\}$ admits continuous
parametrization $s\in\fR$, satisfying
$\lim_{s\searrow 0}\xsym^s=\id$,
$\xsym^s\circ\xsym^t=\xsym^{s+t}$,
then one can construct a piece-wise autonomous ({\ie} autonomous on open
intervals of the evolution parameter) vector field
\[
  \frac{\dif}{\dif{s}}\xsym^s(\x)\Big\vert_{s=0} = \xS(\x),
  \qquad
  \oxS
   := \xS^\idi\partial_\idi        
    + \bar{\xS}^\idi\partial_\ibi, 
\]
where barred indices refer to complex-conjugated variables if $\spX$ is a complex
manifold -- real case is obtained by restriction to vanishing imaginary parts.
Differentiation of the symmetry condition
\[
  \xsym^s\circ\greX \cong \greX\circ\xsym^s
\]
leads to the tangent map (push-forward)
$\oxS|_\greX = \greX_*\oxS$,
which in local coordinates has the form
\begin{equation}
 \label{eq:pushfwd}
  \xS^\idi\circ\greX = \greX^\idi\null\!_{,\idj}\xS^\idj,
\end{equation}
provided $\greX$ is holomorphic.
As pointed out by Maeda \cite{bb:Mae87} and Quispel {\etal}
\cite{bb:QS93,bb:SBQ96}, existence of the vector field $\phi$
satisfying \eqref{eq:pushfwd}, can be used
to reduce dimensionality of Eq. \eqref{eq:OdE} in the same way as
it is done in the theory of differential equations.
More precisely, if there exists an {\IPM} $\greX^t$ such that the symmetry
condition holds true for all $t\in\fR_+$, then the corresponding vector field
$\oX$ defined by
\[
  \frac{\dif}{\dif{t}}\greX^t(\x)\Big\vert_{t=0} = \X(\x),
  \qquad
  \oX
   := \X^\idi\partial_\idi 
    + \bar{\X}^\idi\partial_\ibi,
\]
satisfies the \term{Lie point-symmetry} condition
\[
  [\oX,\oxS]=0.
\]

\section{One-parameter family of similarities}
\label{sc:method}
  Instead of constructing symmetries and similarities upon linearizations, we
now consider a situation when one knows at least some of them in advance. The
question arises as to whether this information can be exploited to determine the
linearizations, and in this paragraph we investigate this possibility.

  The generating function $f$ together with its symmetries form a semigroup
with respect to composition, equipped with unity. That the function in concern
$\greX$ may be part of a structured set of maps, is a key element, which
significantly simplifies the search for a homomorphism with linear
representation of the same structure, that is -- a linearization.

  Similarity between two maps generalizes the notion of a symmetry.
Analogously, a set of similarities form a semigroup with respect to composition
\[
  \begin{aligned}
  \autX:=\{&\xaut(\iga):\spX\rightrightarrows\spX\mid\\
    & \forall_{\iga,\igb}\,\exists_\igc:\;
      \xaut(\iga)\circ\xaut(\igb)\cong\xaut(\igc)\circ\xaut(\iga)\}.
  \end{aligned}
\]
In particular, if all elements of $\autX$ are inverible then this set becomes an
inner automorphism group with composition
$
  \xaut(\iga):\xaut(\igb)
    \mapsto\xaut(\iga)\circ\xaut(\igb)\circ\xaut^{-1}(\iga)
    = \xaut(\igc)
$.

  Suppose, that $\autX$ is continuously parametrized, and contains the function
$\greX$. The structure of this set can be used to determine (formally) isomorphic
semigroup $\autY=\{\ysim(\iga)\}$ of linear maps acting on some vector space
$\spE$. Assuming there exists a homomorphism between $\autX$ and $\autY$ the
problem of linearization reduces to finding a map $\mpp:\spE\to\spX$ such that
\begin{equation}
  \label{eq:grYpX}
  \forall_\iga:\;
  \mpp\circ\ysim(\iga) = \xaut(\iga)\circ\mpp.
\end{equation}
Let the explicit, coordinate form of $\ysim(\iga)$ be
\[
  \ysim^\ida(\pty;\iga)=\mL^\ida\null_\idb(\iga)\y^\idb,
  \qquad
  \begin{aligned}
    \null&\ida,\idb=1\Till\dimY,\\
    \null&\mL^\ida\null_\idb(\iga)\in\fC.
  \end{aligned}
\]
Differentiating with respect to $\pty$ and $\iga$ gives two independent
relations
\begin{align*}
  \mpp^\idi\null\!_{,\ida}(\mL\vey)\mL^\ida\null_{\idb}
    &= \xaut^\idi\null\!_{,\idj}(\mpp(\pty))\mpp^\idj\null\!_{,\idb}(\pty)\\
  \mpp^\idi\null\!_{,\ida}(\mL\vey)\partial_\iga\mL^\ida\null_{\idb}\y^\idb
    &= \partial_\iga\xaut^\idi(\mpp(\pty))
\end{align*}
where $\vey=\{\y^\ida\}$ is the coordinate vector of the point $\pty$,
$\partial_\iga\equiv\partial/\partial\iga$.
Comparing the above equations we get
\begin{equation}
  \label{eq:mppPDE}
  \xaut^\idi\null\!_{,\idj}\mpp^\idj\null\!_{,\idb}
  \partial_\iga\mE^\idb\null_{\idc}\y^\idc
    = \partial_\iga\xaut^\idi.
\end{equation}
where all omitted argumets of $\mpp$ are $\pty$, and as before
$
  \mL(\iga)=\e{\mE(\iga)}
$, that is
$
  \partial_\iga\mE^\idb\null_{\idc}
    = [\mL^{-1}]^\idb\null_{\idd}\partial_\iga\mL^\idd\null_\idc
$.
Therefore, the existence of continuous $1$-parameter semigroup of similarities
leads to the first order system of partial differential equations ({\PDE}s) for
the mapping $\mpp$. Its solution, together with the linear map $\mL(\iga)$ such
that $\xaut(\iga)=\greX$, provides the desired linearization of the system
\eqref{eq:OdE}.

We stress, that the crucial point in the above construction is the formal
isomorphism between structured sets of nonlinear and linear mappings, and not
their particular composition rule -- {\cf} example \ref{sc:Riccati} below.

\subsection{Example: homographic map}
\label{sc:Riccati}
For an easy illustration of the procedure described above, we first analyse the
case of Riccati equation. Let $\x\in\chX=\fC$, and the dynamical system
\eqref{eq:OdE} be the equation generated by a homography $\greX:\chX\to\chX$
\begin{equation}
 \label{eq:Riccati}
  \x_{n+1}
    = \greX(\x_n):=\frac{\iga\x_n+\igb}{\igc\x_n+\igd},
  \qquad
  \iga\igd\not=\igc\igb.
\end{equation}
Recall, that the set of all homographies is a Kleinian group, and under the
above constraint imposed on the parameters, which exclude uninteresting
constant case, it becomes isomorphic with $\grSL(2,\fC)$. A local
parametrization can be given by
\begin{subequations}
 \label{eqs:HRiccati}
 \begin{align}
  \autX = &\left\{
    \xaut(\x)=\frac{\iga(\igd+\igc\x)-1}{\igc(\igd+\igc\x)},\;\igd\not=0\right\}
  \label{eq:HRiccatia}\\
  \autY = &\left\{\mL =\begin{pmatrix} \iga & \frac{\iga\igd-1}{\igc}\\\igc&\igd\end{pmatrix},\;\igd\not=0\right\}
  \label{eq:HRiccatib}
 \end{align}
\end{subequations}
Clearly $\greX\in\autX$.

Out of the three continuous parameters $(\iga,\igc,\igd)$ it is sufficient to
take the one specific -- choose $\iga$ for concreteness. The differential
constraint \eqref{eq:mppPDE} becomes
\[
  \frac{\partial\xaut}{\partial\x}\frac{\partial\mpp}{\partial\y^\idb}
  [\mL^{-1}]^\idb\null_\idc\frac{\partial}{\partial\iga}\mL^\idc\null_{\idd}\y^\idd
    = \frac{\partial\xaut}{\partial\iga}.
\]
Inserting \eqref{eqs:HRiccati}, one gets single {\PDE} of the form
\[
  \frac{\partial\mpp}{\partial\y^1}\frac{\igd}{(\igd+\igc\mpp)^2}
 -\frac{\partial\mpp}{\partial\y^2}\frac{\igc}{(\igd+\igc\mpp)^2}
 =\frac{1}{\igc\y^1+\igd\y^2}.
\]
The solution of this equation has two integration constants $C\not=0$ and $D$:
\begin{equation}
 \label{eq:mppsol}
  \mpp(\pty)=\frac{\y^1+\igd C D}{\y^2-\igc C D}.
\end{equation}
Taking into account \eqref{eq:grYpX}, \eqref{eqs:HRiccati} and \eqref{eq:mppsol},
one finds that the only possibility left to satisfy $C\not=0$ is to set $D=0$.
Therefore
\[
  \mpp(\pty)=\frac{\y^1}{\y^2},
\]
which is the desired linearizing map.

  Having found the linearization, it is straightforward to provide the inverse
{\Poincare} maps $\greX^t(\x)$ of the Riccati equation \eqref{eq:Riccati}. These
maps are also homographies with $t$-dependent coefficients, and in general, for
arbitrary initial position $\x\in\chX$ they go to infinity in finite time. This
implies, that the motion is better described in complex projective space
$\spX=\fC\cup\{\infty\}=\fC P^1$, instead of a single chart $\chX$. Taking the
second local chart as $\chX'=\fC$, with transition functions $\mpt:\chX\to\chX'$
$
  \x' = \mpt(\x) := 1/\x,
$
one has $\greX'=\mpt\circ\greX\circ\mpt^{-1}$, that is
\[
  \x'_{n+1}
    = \greX'(\x'_n)
    = \frac{\igc(\igd\x'_n + \igc)}{(\iga\igd - 1)\x'_n+\iga\igc}.
\]
The projection map in in these coordinates $\mpp'=\mpt\circ\mpp$, is simply
$
  \mpp'(\pty)={\y^2}/{\y^1}
$.
Both $\mpp$ and $\mpp'$ define local inhomogeneous coordinates of the
projective space $\spX=\fC P^1$.

\subsection{Example: the logistic equation}
Consider a more involved example of the logistic equation
\begin{equation}
  \label{eq:Log4}
  \x_{n+1}
    = \greX(\x_n) := 4\x_n(1-\x_n).
\end{equation}
It is well-known that it is linearizable -- we now recover this linearization
following the approach described earlier. We first observe, that $\greX$ is a
member of the following set\footnote{
 The branches of infinitely multivalued map $\acos:[-1,1]\rightrightarrows\fR$,
 can be expressed explicitly as
 $\acos_k(x)=(-1)^k\Acos(x)+\pi[k-((-1)^k-1)/2]$,
 where $\Acos:[-1,1]\to[0,\pi]$ is the principal branch.
}
\begin{equation}
 \label{eq:HLogistic}
  \hspace{-1mm}
  \autX
    = \big\{\xaut(\x;\iga)=\sin^2\left[\pi\iga+\acos(1-2\x)\right]\mid
            \iga\in\fR\big\}
\end{equation}
for $\iga\in\fZ$. The compositions
\[
  \xaut(\iga)\circ\xaut(\igb)\cong\xaut(\igc)\circ\xaut(\iga),
\]
satisfy either
\[
  \igc=2\igb-\iga
  \qquad\text{or}\qquad
  \igc=3\iga-2\igb.
\]
The following linear operators $\mL_\pm=\mL_\pm(\iga)$
\[
  \mL_{+}=\begin{bmatrix}2&\iga\\0&1\end{bmatrix},
  \quad\text{and}\quad
  \mL_{-}=\begin{bmatrix}-2&\iga\\0&1\end{bmatrix}.
\]
define two representations $\autY$ of \eqref{eq:HLogistic}, that is
\begin{align*}
  \mL_+(\iga)\mL_+(\igb)&=\mL_+(2\igb-\iga) \mL_+(\iga),\\
  \mL_-(\iga)\mL_-(\igb)&=\mL_-(3\iga-2\igb)\mL_-(\iga).
\end{align*}
The equations \eqref{eq:mppPDE}, in each case reduce to a single {\PDE}
\[
  -\frac{\partial\mpp}{\partial\y^1}=\pm\frac{2\pi}{\y^2}\sqrt{\mpp(1-\mpp)},
\]
where the `$\pm$' signs correspond to $\mL_{+}$ and $\mL_{-}$ respectively.
Their solutions however, are formally identical
\[
  \mpp(\pty)=\thalf-\thalf\cos\left(\pi\tfrac{\y^1}{\y^2}+C_\pm\right),
  \qquad
  C_\pm=C_\pm(\y^2).
\]
Substituting $\mpp(\pty)$ into Eq. \eqref{eq:grYpX}
$
  \mpp(\mL_{\pm}\vey)=\xaut(\mpp(x))
$,
we find the functions $C_\pm$ to be constants, and for each label `$\pm$', two
further possibilities arise, namely we have
\[
  \begin{aligned}
    \mpp_{+,1}(\pty)&=\thalf-\thalf\cos\big(\pi\big(\tfrac{\y^1}{\y^2}&-            &\,\iga\big)\big),\\
    \mpp_{+,2}(\pty)&=\thalf-\thalf\cos\big(\pi\big(\tfrac{\y^1}{\y^2}&+3           &\,\iga\big)\big),\\
    \mpp_{-,1}(\pty)&=\thalf-\thalf\cos\big(\pi\big(\tfrac{\y^1}{\y^2}&-\tfrac{1}{3}&\,\iga\big)\big),\\
    \mpp_{-,2}(\pty)&=\thalf-\thalf\cos\big(\pi\big(\tfrac{\y^1}{\y^2}&+            &\,\iga\big)\big).
  \end{aligned}
\]
The corresponding inverses are
\[
  \begin{aligned}
    \mps_{+,1}(\x)&=\big[\y^2\big(\tfrac{1}{\pi}\acos(1-2\x)&+            &\,\iga\big),\y^2\big],\\
    \mps_{+,2}(\x)&=\big[\y^2\big(\tfrac{1}{\pi}\acos(1-2\x)&-3           &\,\iga\big),\y^2\big],\\
    \mps_{-,1}(\x)&=\big[\y^2\big(\tfrac{1}{\pi}\acos(1-2\x)&+\tfrac{1}{3}&\,\iga\big),\y^2\big],\\
    \mps_{-,2}(\x)&=\big[\y^2\big(\tfrac{1}{\pi}\acos(1-2\x)&-            &\,\iga\big),\y^2\big].
  \end{aligned}
\]
The nonlinear {\IPM}s can be computed from
\begin{equation}
  \label{eq:IPMsol}
  \greX^t_\pm=\mpp\circ\mL^t_\pm\circ\mps
\end{equation}
where the powers of $\mL^t_\pm$ are\footnote{
 In the complex domain both $2^t$ and $(-2)^t$ are countably many-valued. A
 particular branch $\mE_\pm=\ln\mL_\pm$ is selected by fixing the integer $k$
 in $(\pm2)^t=\exp[t\ln 2 +t\imu\pi(2k+(1\mp 1)/2)]$.
}
\[
  \begin{aligned}
  \mL_{+}^t&=\begin{bmatrix}2^t&\iga(2^t-1)\\0&1\end{bmatrix},\\
  \mL_{-}^t&=\begin{bmatrix}(-2)^t&\frac{\iga}{3}(1-(-2)^t)\\0&1\end{bmatrix}.
  \end{aligned}
\]
Now, the composition \eqref{eq:IPMsol} yields
\[
  \hspace{-8mm}
  \begin{aligned}
    \greX_{+,1}^t(\x)=\sin^2\big(&2^{t-1}\acos(1-2\x)+\pi\iga(2^t-1)\big),\\
    \greX_{+,2}^t(\x)=\sin^2\big(&2^{t-1}\acos(1-2\x)+\pi\iga(1-2^t)\big),\\
    \greX_{-,1}^t(\x)=\sin^2\big(&(-2)^{t-1}\acos(1-2\x)\\
                                +&\tfrac{2\pi}{3}\iga(1-(-2)^t)\big),\\
    \greX_{-,2}^t(\x)=\sin^2\big(&(-2)^{t-1}\acos(1-2\x)]\big).
  \end{aligned}
\]
Clearly, $\greX_{+,1}^t$ and $\greX_{+,2}^t$ represent the same family of maps,
and for the second pair, $\greX_{-,2}^t$ is a subset of $\greX_{-,1}^t$.
Therefore, there are actually two distinct sets of (multivalued)
continuous trajectories $\greX_{+}^t$ and $\greX_{-}^t$, however only in the
first one there exists a real-valued branch of $2^t$, and hence
$\greX_{+}^t(\x)\in\fR$, provided $\x\in[0,1]$. All remaining trajectories are
complex.

\begin{figure}
 \includegraphics[trim=0 540 0 0,clip=true]{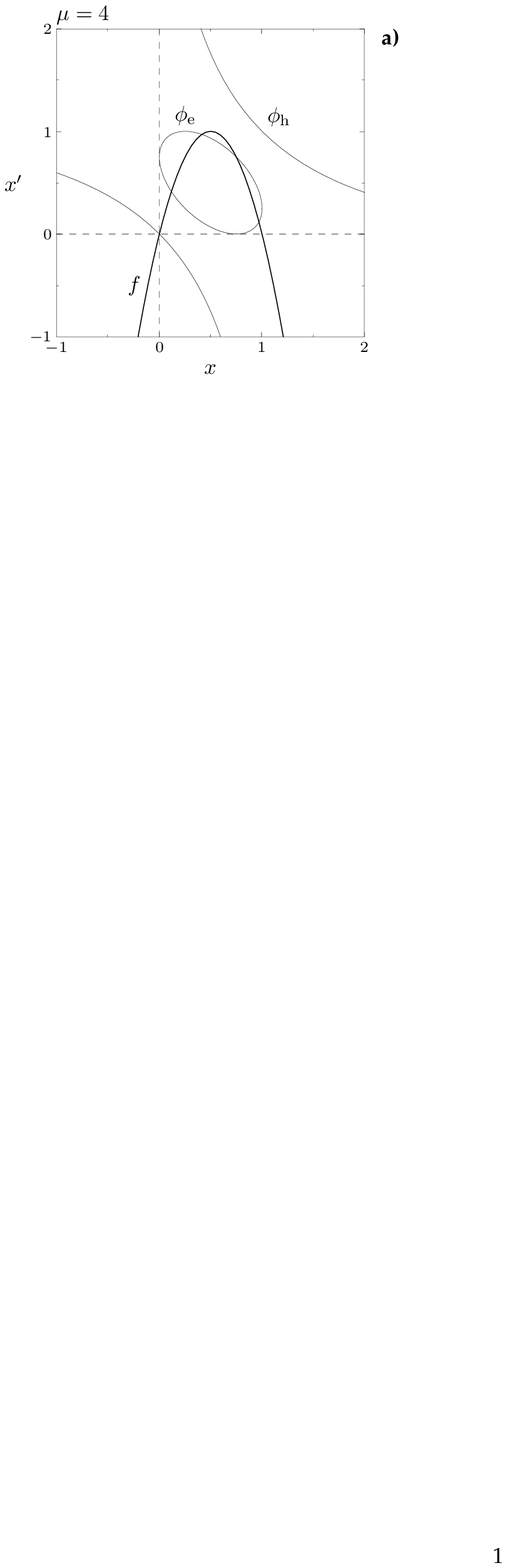}\par\indent
 \includegraphics[trim=0 540 0 0,clip=true]{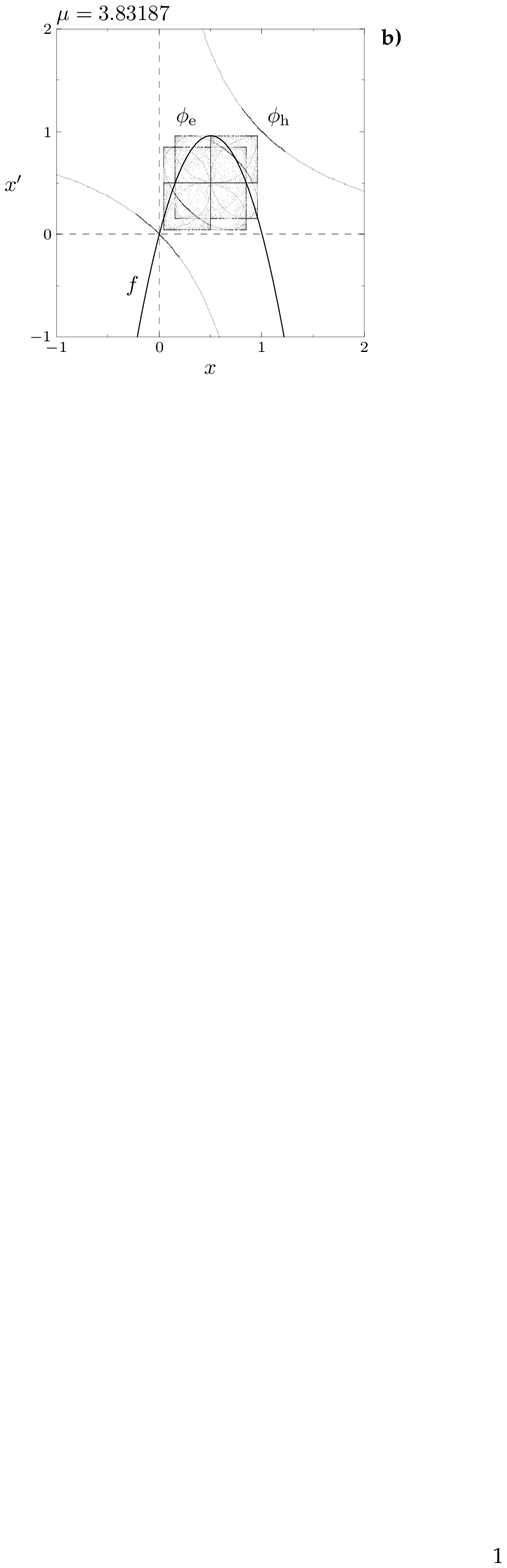}\par\indent
 \includegraphics[trim=0 540 0 0,clip=true]{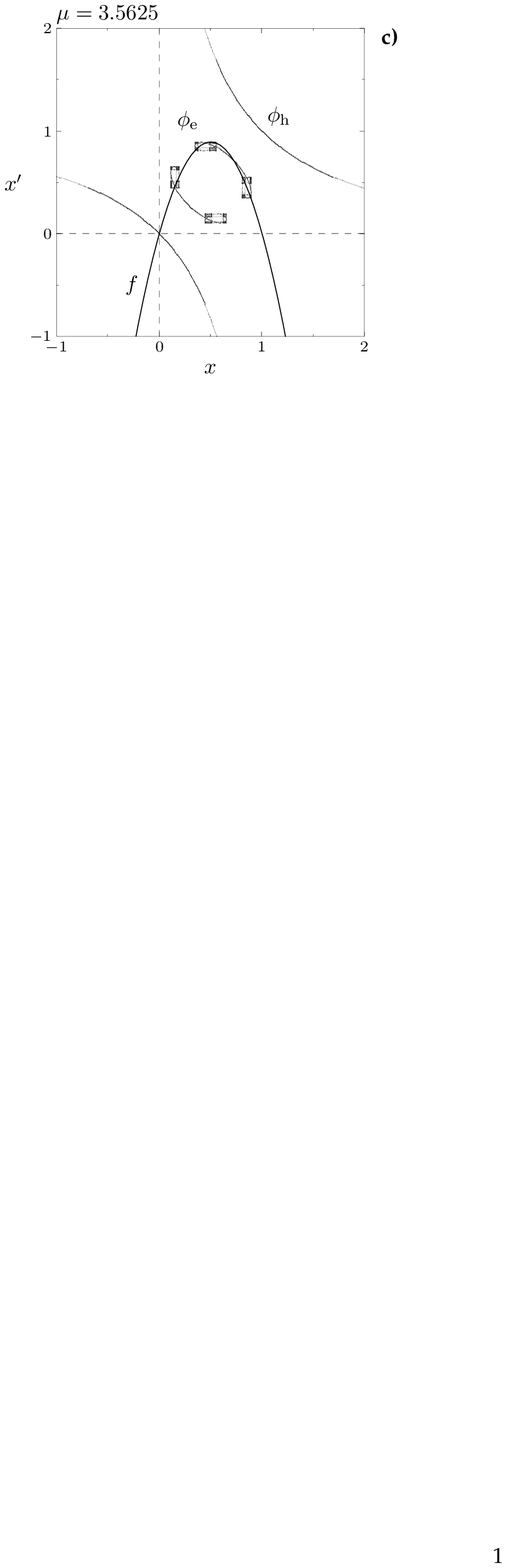}\par\indent
 \includegraphics[trim=0 540 0 0,clip=true]{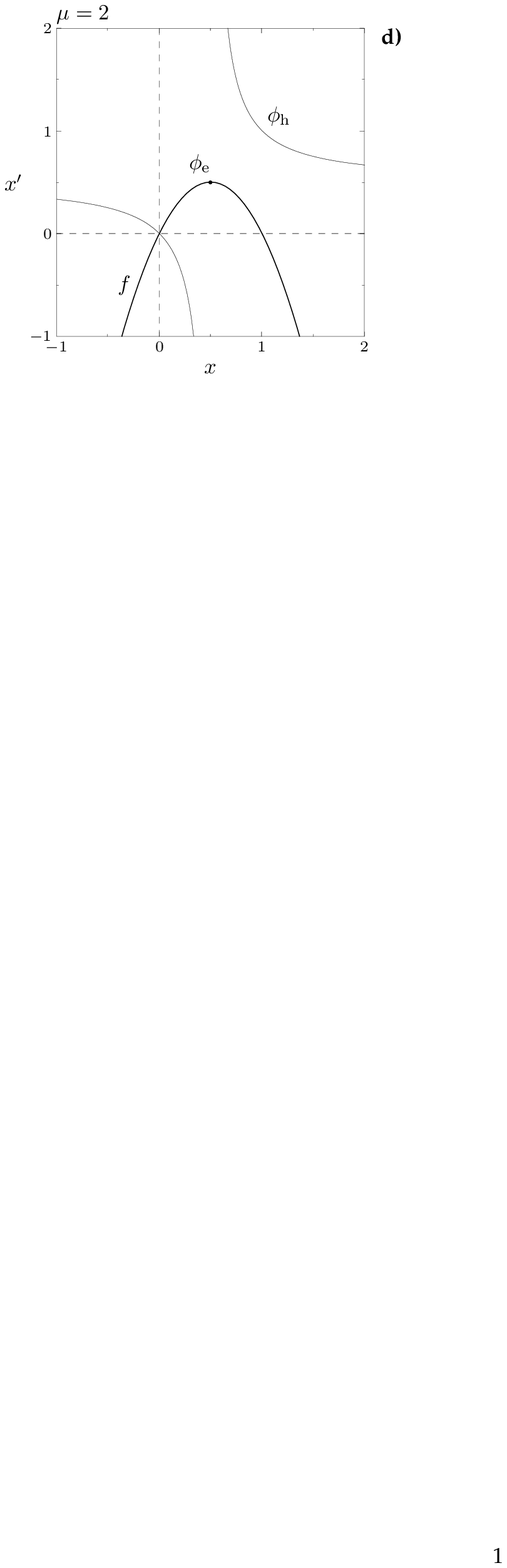}
 \figcaption{Examples of the symmetry curves of general logistic equation
  $\greX(\x)=\mu\x(1-\x)$, in several cases of its control parameter
  $\mu$. Due to $\greX\circ s=\greX$, $s(x)=1-x$, both $\phi_\mathrm{e}$ and
  $\phi_\mathrm{h}$ have point-reflection isometry through
  $(x,x')=(\thalf,\thalf)$. The two cases $\mu=3.83187$ (close to periodic
  window) and $\mu=3.5625$ (just before onset of chaos), were
  found numerically through procedure described in the text. The curves
  $\phi_\mathrm{e}$ on \textbf{b)}, \textbf{c)} seem to be fractal, for not
  being closed and filling densely subsets of $\tilde{\spX}^2=\fR^2$.}
 \label{fig:LogSym}
\end{figure}

\medskip
Our purpose now, is to discuss the symmetries constructed from obtained
linearizations. Since $\mL_{\pm,1}=\mL_{\pm,2}$, therefore we use $U=\idmatrix$
as a solution to \eqref{eq:EUUE}. The first pair composed according to Eq.
\eqref{eq:xsim}, that is
\begin{align*}
  \mpp_{+,1}\circ\mps_{+,2}
    &\cong \mpp_{+,2}\circ\mps_{+,1}\\
    &\cong \sin^2\left[\pi\iga+\thalf\acos(1-2\x)\right],
\end{align*}
becomes an identity for $\iga\in\fZ$, which is a trivial symmetry.
But the second pair
\begin{align*}
  \mpp_{-,1}\circ\mps_{-,2}
    &\cong \mpp_{-,2}\circ\mps_{-,1}\\
    &\cong \sin^2\left[\tfrac{2}{3}\pi\iga+\thalf\acos(1-2\x)\right]\\
    &=:\xsym_\mathrm{e}(\x)
\end{align*}
yield a nontrivial one at $\iga=1,2+3k$, $k\in\fZ$. After reduction this is
\begin{subequations}
 \begin{equation}
 \label{eq:xsyme}
  \xsym_\mathrm{e}(\x)=\tfrac{1}{4}\left(3-2\x\pm 2\sqrt{3\x(1-\x)}\right),
 \end{equation}
namely, the solution $\x'=\xsym_\mathrm{e}(\x)$ to an equation of ellipse
inscribed into the square $[0,1]\times[0,1]$:
 \begin{equation}
 \label{eq:xSyme}
  \xSym_\mathrm{e}(x,x'):=
    (\x+\x'-1)^2+\tfrac{1}{3}(\x-\x')^2-\tfrac{1}{4}=0.
 \end{equation}
\end{subequations}
Compositions of $\xsym_\mathrm{e}$ with $\greX^t_\pm$ gives uncountable number
of symmetries out of this one. Nevertheless, this set is not exchaustive; other
symmetries can for instance be constructed according to Eq. \eqref{eq:PhaseSym}.
In order to illustrate, consider the following linear maps
\[
  \mM_+:=\begin{bmatrix}\imu & (\imu-1)\iga \\ 0 & 1 \end{bmatrix},
  \qquad
  \mM_-:=\begin{bmatrix}\imu & \tfrac{1}{3}(1-\imu)\iga \\ 0 & 1 \end{bmatrix},
\]
trivially satisfying the condition \eqref{eq:ComtKer}, by $[\mM_\pm,\mL_\pm]=0$.
Restricting to the case of integer parameter $\iga$,
the compositions $\mpp_\pm\circ\mM_\pm\circ\mps_\pm=:\xsym_{\pm,\mathrm{h}}$
take the form
\begin{align*}
  \xsym_{+,\mathrm{h}}
    &=-\sinh^2\big[\pi\iga+\thalf\acos(1-2\x)\big],\\
  \xsym_{-,\mathrm{h}}
    &=-\sinh^2\big[\tfrac{2\pi}{3}(1+\imu)\iga+\thalf\acos(1-2\x)\big].
\end{align*}
We point out that, neither of these map the interval $[0,1]$ into itself, as
$\xsym_\mathrm{e}$ does, therefore they are not proper symmetries of the
logistic equation in the sense of definition given earlier.
In particular, both the above sets have one common element for
$\iga=0$. It is rather special one for being real-valued and self-inverse
$\xsym_\mathrm{h}^{-1}=\xsym_\mathrm{h}$.
\[
  \xsym_\mathrm{h}:\fR\rightrightarrows\fR,
\]
\[
  \xsym_\mathrm{h}: x\mapsto
    \begin{cases}
      \sin ^2\big[\thalf\acosh(1-2 x)\big] & x\in(-\infty,0],\\
    \begin{matrix}
     -\sinh^2\big[\thalf\acos (1-2 x)\big]\\
      \cosh^2\big[\thalf\acos (2 x-1)\big]
    \end{matrix}
     & x\in[0,1],\\
      \cos ^2\big[\thalf\acosh(2 x-1)\big] & x\in[1,\infty)
    \end{cases}
\]
This relation $\xsym_\mathrm{h}$, together with $\xsym_\mathrm{e}$, are
illustrated in Fig.~\ref{fig:LogSym}a).

  We have also investigated symmetries of the logistic equation
$\x_{n+1}=\mu\x_n(1-\x_n)$ numerically, in other cases of the parameter $\mu$.
The applied procedure directly uses the symmetry
definition \eqref{eq:defSym}. Namely, assuming $\xsym(x_*)=x_*$, one can
develop $\xsym$ into power series \eqref{eq:symexp} in some neighbourhood of
the fixed point $x_*$, as described in Sect. \ref{sc:Sym}. We confine ourselves
to the self-inverse symmetries $\xsym^{-1}=\xsym$, by fixing the free expansion
parameter to $\partial_x\xsym(x_*)=-1$. Starting from $x_0:=x_*+\delta{x}$,
$|\delta{x}|\ll 1$, the expansion can then be used to approximate
$x_0':=\xsym(\x_0)$ to an arbitrary precision. By iteration
\[
  \{x_n,x'_n\}\mapsto\{x_{n+1},x'_{n+1}\}=\{f(x_n),f(x'_n)\},
\]
a finite number of points belonging to the symmetry relation is obtained.
Many repetitions of this algorithm, with various initial displacements
$\delta{x}$ yielded sufficient data to draw reliable pictures of $\xsym$ in
number of cases $\mu\in[2,4]$ -- two particular are shown on Fig.~\ref{fig:LogSym}b,c)

  Both $\xsym_\mathrm{e}$ and $\xsym_\mathrm{h}$ exist for all checked
values of the control parameter $\mu$, except $\mu=2$ where $\xsym_\mathrm{e}$
degenerates to a single point at $(x_*,x_*)=(\thalf,\thalf)$. At this extreme
the second curve $\xsym_\mathrm{h}$ can be given closed analytical form, thanks
to the well-known linearization
\begin{align*}
  &2 x(1-x)=\mpp(2\mps(x)),\\
  &\mpp(y)=\thalf\big(1-\e{2 y}\big),\quad
  \mps(x)=\thalf\ln(1-2 x).
\end{align*}
Again, taking $\mM=-1$ in Eq. \eqref{eq:PhaseSym} gives
({\cf} Fig.~\ref{fig:LogSym}d))
\begin{subequations}
 \begin{equation}
 \label{eq:xsymh}
  \xsym_\mathrm{h}(x) = \mpp(-\mps(x))=\frac{x}{2 x - 1},
 \end{equation}
or, in the equivalent form ({\cf} Eqs. \eqref{eq:xsyme}, \eqref{eq:xSyme}),
the solution $\x'=\xsym_\mathrm{h}(\x)$ to
 \begin{equation}
 \label{eq:xSymh}
  \xSym_\mathrm{h}(x,x')=x x'-\thalf(x+x')=0.
 \end{equation}
\end{subequations}
It should be noted that, in contrast to all other cases of $\mu>2$,
here $\xsym_\mathrm{h}$ is an ordinary ({\ie} single-valued) function.

We also remark, that by construction each of the fixed points $\x_*=0,1-\mu^{-1}$
of the vector field $f$ is also a fixed point of one of the two symmetries
$\xsym_\mathrm{h}$ and $\xsym_\mathrm{e}$ respectively:
\begin{align*}
  0          &\not\in\img\phi_\mathrm{e}(0),&\quad
  1-\mu^{-1} &\in\img\phi_\mathrm{e}(1-\mu^{-1}),\\
  0          &\in\img\phi_\mathrm{h}(0),&\quad
  1-\mu^{-1} &\not\in\img\phi_\mathrm{h}(1-\mu^{-1})
\end{align*}
This is not generally the case for the time-translation symmetries $f^t_\pm$.
It is therefore plausible to conjecture that, apart from special degenerate
cases, presence of non-trivial discrete symmetries associated with equilibrium
points of the vector field $f:\spX\to\spX$ is a generic feature in difference
equations.
In this sense, $\xsym_\mathrm{e}$ and $\xsym_\mathrm{h}$ can be considered
typical of logistic map $f(x)=\mu\x(1-\x)$.

\section{Conclusions}
In this paper, we have investigated the relationships among symmetries,
similarities, linearizations and inverse {\Poincare} maps of ordinary difference
equations.
  The general conclusion is, that the existence of additional, continuously
parametrized family of similarities, admissible by the system, plays an
important role for the analytic solution of {\OdE} and/or its linearization.
Existence of such structure can be regarded as an analog of Lie algebra in
the case of Lie-type ordinary differential systems.
This structure generalizes the notion of a symmetry for it does not necessarily
have to be a group, but may instead form a semigroup without unity. The
important point is, that if a linear representation of this structure exists,
then the linearization problem of original {\OdE} can be reduced to a solution
of a finite partial differential system.

  Finally, as suggested by the results presented here, the number of non-trivial
discrete symmetries appears to be directly related to the cardinality of the
fixed subset of phase space $\spX$. Interestingly, these relations need not to
be simultaneously symmetries of the inverse {\Poincare} maps $f^t$, but
instead provide similarity transformations among them.


\def\bbaut#1#2{{#2} {#1},}
\def\bbttl#1{\textsl{``#1''},}
\def\bbpub#1{\textit{#1}}
\def\bbepr#1{\texttt{#1}}

\clearpage
\end{multicols}

\begin{thebibliography}{AAA}{\footnotesize
\setlength{\itemsep}{0pt}\setlength{\parskip}{0pt}

\bibitem{bb:SR75}
  \bbaut{M.}{Reed}
  \bbaut{B.}{Simon}
  \bbttl{Methods of modern mathematical physics II: Fourier analysis, self-adjointness}
  Academic Press, New York 1975.

\bibitem{bb:Mae87}
  \bbaut{S.}{Maeda}
  \bbttl{The similarity method for difference equations}
  \bbpub{IMA J. Appl. Math.} \textbf{38} (1987), 129--134.

\bibitem{bb:QS93}
  \bbaut{G.R.W.}{Quispel}
  \bbaut{R.}{Sahadevan}
  \bbttl{Lie symmetries and the integration of difference equations}
  \bbpub{Phys. Lett. A} \textbf{184} (1993), 64--69.

\bibitem{bb:SBQ95}
  \bbaut{G.B.}{Byrnes}
  \bbaut{R.}{Sahadevan}
  \bbaut{G.R.W.}{Quispel}
  \bbttl{Factorizable Lie symmetries and the linearization of difference equations}
  \bbpub{Nonlinearity} \textbf{8} (1995) 443--459.

\bibitem{bb:SBQ96}
  \bbaut{R.}{Sahadevan}
  \bbaut{G.B.}{Byrnes}
  \bbaut{G.R.W.}{Quispel}
  \bbttl{Linearization of difference equations using factorisable Lie symmetries}
  in \bbpub{Symmetries and integrability of difference equations}
  ed. D. Levi, L. Vinet, P. Winternitz, CRM Proc \& Lecture Notes \textbf{8} AMS,
  1996, pp 337--343;

\bibitem{bb:Olv86}
  \bbaut{P.J.}{Olver}
  \bbttl{Applications of Lie groups to differential equations}
  Springer-Verlag, New York, 1996.

\bibitem{bb:BlKu89}
  \bbaut{G.W.}{Bluman}
  \bbaut{S.}{Kumei}
  \bbttl{Symmetries and differential equations}
  Springer-Verlag, New York, 2nd. ed. 1989.

\bibitem{bb:Stp89}
  \bbaut{H.}{Stephani}
  \bbttl{Differential equations. Their solution using symmetries}
  Cambridge Univ. Press, Cambridge 1989.

}\end{thebibliography}
\end{document}